\documentclass[lettersize,journal]{IEEEtran}

\usepackage{amsmath,amsthm,amsfonts}
\usepackage{tablefootnote}
\usepackage[Algorithm]{algorithm}
\usepackage{algpseudocode}
\usepackage{array}
\usepackage[caption=false,font=normalsize,labelfont=sf,textfont=sf]{subfig}
\usepackage{textcomp}
\usepackage{stfloats}
\usepackage{url}
\usepackage{verbatim}
\usepackage{xcolor}
\usepackage{stmaryrd}
\usepackage{graphicx}
\usepackage{multirow}
\usepackage{cite}
\usepackage{siunitx}
\usepackage{threeparttable}
\usepackage[shortlabels]{enumitem} %
\usepackage{csquotes}
\usepackage{pifont}%
\newcommand{\cmark}{\ding{51}}%
\newcommand{\xmark}{\ding{55}}%
\usepackage{array, tabularx, ragged2e, makecell, booktabs, longtable} %
\renewcommand{\theadfont}{\normalsize} %
\renewcommand{\arraystretch}{1.15}
\newcolumntype{L}{>{\RaggedRight\arraybackslash\hspace{0pt}}X} %
\newcolumntype{R}{>{\RaggedLeft\arraybackslash\hspace{0pt}}X} %
\usepackage{adjustbox}
\usepackage{scalerel}
\usepackage{tikz}
\usetikzlibrary{svg.path}
\definecolor{orcidlogocol}{HTML}{A6CE39}
\tikzset{
  orcidlogo/.pic={
    \fill[orcidlogocol] svg{M256,128c0,70.7-57.3,128-128,128C57.3,256,0,198.7,0,128C0,57.3,57.3,0,128,0C198.7,0,256,57.3,256,128z};
    \fill[white] svg{M86.3,186.2H70.9V79.1h15.4v48.4V186.2z}
                 svg{M108.9,79.1h41.6c39.6,0,57,28.3,57,53.6c0,27.5-21.5,53.6-56.8,53.6h-41.8V79.1z M124.3,172.4h24.5c34.9,0,42.9-26.5,42.9-39.7c0-21.5-13.7-39.7-43.7-39.7h-23.7V172.4z}
                 svg{M88.7,56.8c0,5.5-4.5,10.1-10.1,10.1c-5.6,0-10.1-4.6-10.1-10.1c0-5.6,4.5-10.1,10.1-10.1C84.2,46.7,88.7,51.3,88.7,56.8z};
  }
}

\newcommand\orcidicon[1]{\href{https://orcid.org/#1}{\mbox{\scalerel*{
\begin{tikzpicture}[yscale=-1,transform shape]
\pic{orcidlogo};
\end{tikzpicture}
}{|}}}}

\usepackage{hyperref}
\usepackage{cleveref}

\usepackage[
lambda,
advantage,
operators,
sets,
adversary,
landau,
probability,
notions,
logic,
ff,
mm,
primitives,
events,
complexity,
asymptotics,
keys
]{cryptocode}

\usepackage{ulem}
\usepackage{lipsum}
\usepackage{xspace}
\newcommand{\mathcmd}[1]{\ensuremath{#1}\xspace}
\newcommand{\R}{\mathbb{R}} 
\newcommand{\N}{\mathbb{N}} 
\newcommand{\C}{\mathbb{C}}
 
\newcommand{\Z}{\mathbb{Z}}

\renewcommand{\P}{\mathsf{P}}

\newcommand{\ucf}{\mathcmd{\mathcal{F}}}
\newcommand{\ucz}{\mathcmd{\mathcal{Z}}}
\newcommand{\ucs}{\mathcmd{\mathcal{S}}}
\newcommand{\ucr}{\mathcmd{\mathcal{R}}}

\newcommand{\ab}{\boldsymbol{a}}
\newcommand{\bb}{\boldsymbol{b}}

\newcommand{\yb}{\boldsymbol{y}}
\newcommand{\ib}{\boldsymbol{i}}

\newcommand{\gb}{\boldsymbol{g}}

\newcommand{\wb}{\boldsymbol{w}}

\newcommand{\xb}{\boldsymbol{x}}

\newcommand{\zb}{\boldsymbol{z}}
\newcommand{\qb}{\boldsymbol{q}}
\newcommand{\pb}{\boldsymbol{p}}
\newcommand{\vb}{\mathbf{v}}
\renewcommand{\sb}{\boldsymbol{s}}
\newcommand{\vs}{\mathrm{v}}

\newcommand{\Ab}{\boldsymbol{A}}
\newcommand{\Bb}{\boldsymbol{B}}

\newcommand{\Hb}{\boldsymbol{H}}
\newcommand{\Lb}{\boldsymbol{L}}

\newcommand{\Fb}{\boldsymbol{F}}
\newcommand{\Gb}{\boldsymbol{G}}
\newcommand{\Ib}{\boldsymbol{I}}
\newcommand{\Jb}{\boldsymbol{J}}

\newcommand{\Pb}{\boldsymbol{P}}

\newcommand{\Yb}{\boldsymbol{Y}}
\newcommand{\Wb}{\boldsymbol{W}}

\newcommand{\Kb}{\boldsymbol{K}}
\newcommand{\Ub}{\boldsymbol{U}}
\newcommand{\Vb}{\boldsymbol{V}}

\newcommand{\Ps}{\mathsf{P}}

\renewcommand{\Re}[1]{\mathrm{Re}\left(#1\right)}
\renewcommand{\Im}[1]{\mathrm{Im}\left(#1\right)}
\newcommand{\jc}{\mathrm{j}}

\newcommand{\zerob}{\boldsymbol{0}}

\newtheorem{thm}{Theorem}

\newtheorem{defn}{Definition}

\newenvironment{complexity}{\textit{Complexity:}}{\hfill}

\newcommand{\share}[1]{[#1]}

\newcommand{\vect}[1]{\mathrm{vec}\left(#1\right)}
\newcommand{\round}[1]{\left\lfloor#1\right\rceil}
\newcommand{\roundtext}[1]{\lfloor#1\rceil}

\DeclareMathOperator*{\diag}{diag}

\newcommand{\blind}[1]{\textcolor{white}{#1}}

\begin{document}

\title{Privacy-Preserving Power Flow Analysis \\
via Secure Multi-Party Computation}

\author{Jonas von der Heyden$^{\star}$ \orcidicon{0000-0002-2846-5163}, Nils Schlüter$^{\star}$ \orcidicon{0000-0002-0166-199X}, Philipp Binfet \orcidicon{0009-0005-3486-2690}, Martin Asman \orcidicon{0009-0007-0823-5786}, Markus Zdrallek, Tibor Jager, and Moritz Schulze Darup \orcidicon{0000-0002-1868-4098} %
\thanks{
This work was supported in part by the German Research Foundation (DFG) und the grants JA 2445/7-1, SCHU 2940/4-1, SCHU 2940/5-1, and in part by the Daimler and Benz Foundation under grant 32-08/19.}%
\thanks{$^\star$ The first two authors contributed equally to this work.}
\thanks{Jonas von der Heyden, Martin Asman, Tibor Jager and Markus Zdrallek are with the School of Electrical, Information and Media Engineering, Bergische Universität Wuppertal, 42119 Wuppertal, Germany (emails: \{jvdh, asman, jager, zdrallek\}@uni-wuppertal.de).

Nils Schlüter, Philipp Binfet and Moritz Schulze Darup are with the Department of Mechanical Engineering, Technische Universität Dortmund, 44227 Dortmund, Germany (emails: \{nils.schlueter, philipp.binfet, moritz.schulzedarup\}@tu-dortmund.de).}%
\thanks{This version of the manuscript has been accepted for publication in IEEE Transactions on Smart Grid. DOI: \href{https://doi.org/10.1109/TSG.2024.3453491}{10.1109/TSG.2024.3453491}}
}

\IEEEpubid{\copyright~2024 IEEE. Personal use is permitted. For any other purposes, permission must be obtained from IEEE.
}

\maketitle

\begin{abstract}
Smart grids feature %
a bidirectional flow of electricity and data, enhancing %
flexibility, efficiency, and reliability in 
increasingly volatile energy grids.
However, 
data from %
smart meters %
can reveal sensitive private information. 
Consequently, the adoption of smart meters is often restricted via legal means and hampered by limited user acceptance.
Since metering data is beneficial for fault-free grid operation, power management, and resource allocation, applying privacy-preserving techniques to smart metering data is an important research problem.
This work addresses this by using
secure multi-party computation (SMPC), which allows
multiple parties to jointly evaluate functions
of their private inputs without revealing the latter.
Concretely, we show how to perform power flow analysis on cryptographically hidden prosumer data. 
More precisely, we present a tailored solution to the power flow problem building on an SMPC implementation of Newton's method.
We analyze the security of our approach in the universal composability framework and
provide benchmarks for various grid types, threat models, and solvers.
Our results indicate %
that secure multi-party computation can %
be able to alleviate privacy issues in smart grids in certain applications.
\end{abstract}


{
\section*{Nomenclature} 

\begingroup
\renewcommand{\arraystretch}{1} %
\noindent
\begin{tabularx}{\columnwidth}{lX}
    \multicolumn{2}{l}{\textit{Variables}} \\
       \vspace{-3mm}  & \\
    $\mathrm{j}$ & imaginary unit \\
    $N$ & number of buses \\
    $\eta$ & step size \\
    $\vb,\ib$ & vector of voltages, currents \\
    $\pb,\qb,\sb$ & vector of active, reactive, apparent powers \\
    $\xb$ & vector of real and imaginary voltages \\
    $\Vb,\Ib$ & vector of voltage, current magnitudes \\
    $\Fb$ & vector-valued function \\
    $\Bb,\Gb,\Yb$ & susceptance, conductance, admittance matrix \\
    $\Jb$ & Jacobian matrix \\
    $\Pb$ & preconditioner matrix \\
\end{tabularx}

\noindent
\begin{tabularx}{\columnwidth}{lX}
    \vspace{-2.2mm} & \\
    $ \zerob$ & zero vector or matrix  \\
    $ \mathbb{I}$ & Identity matrix \\
    $h$ & exponent of the scaling factor \\
    $n$ & number of parties \\
    $p$ & modulus \\ %
    $r$ & random number \\
    $t$ & number of possibly corrupted parties \\
    $z$ & integer-valued secret \\
    $\lambda$ & security parameter \\
    $\ell$ & maximum iteration number   \\
    $g,\gb$ & cost for step-width, generic vector-valued function \\
    $\yb$ & output vector of generic function \\
    $a,b,c$ & multiplication triples \\
    $\P$ & party in the computation protocol \\
    & \\
\end{tabularx}
\noindent
\begin{tabularx}{1.1\columnwidth}{lX}
 \multicolumn{2}{l}{\textit{Notation}} \\
       \vspace{-3mm}  & \\
    $ \floor{\cdot},\round{\cdot} $ & component-wise flooring, rounding \\
    $a\!\mod p$ & $a$ modulo $p$  \\ %
    $[a]$ & secret-shared number \\
    $[a]^{(i)}$ & $i$th share of a \\
    $\Z_p$ & field of integers modulo $p$ \\ %
    $\Pi$ & Protocol \\
    \ucf, \ucs, \ucz & Functionality, Simulator, Environment \\
    $ a^*$ & complex conjugation \\
    $ \ab \leq \bb $ & component-wise comparison of vectors \\
    $\diag(\ab)$ & zero matrix but with $\ab$ on its diagonal \\
    $ a_i $ & $i$th instance of scalar $a$, $i$th element of vector $\ab$ \\
    $ \ab_k$, $a_{k,i}$ & $k$th instance of vector $\ab$, $i$th element of $\ab_k$ \\
    $  A_{ij} $ & element $ij$ of the matrix $\Ab$ \\
    $ \Ab_{i,:}, \Ab_{:,i}$ & $i$th column, $i$th row of the matrix $\Ab$ \\
\end{tabularx}

\endgroup
\IEEEpubidadjcol %
}

\section{Introduction}
\IEEEPARstart{R}ENEWABLE energy sources play an increasingly important role in addressing energy needs while mitigating environmental impact. Unfortunately, unlike traditional sources, they introduce uncertainty regarding power generation and hence, flexible energy sources (e.g., costly secondary power plants) are currently necessary to ensure balanced loads. Ideally, the power feed-in of renewables %
would not impair the grid's reliability and efficiency.
In this context, power flow analysis (PFA) based on power forecasts can help to contain issues such as voltage instability, overloads, and reactive power requirements \cite{ajjarapu1992continuation} by providing crucial information about the power system's performance.
Here, usage data is a key enabler because it serves as a basis for such forecasts and more reliable planning.
Especially low-voltage networks would benefit from prosumer data since the massive deployment of photovoltaics and batteries has a significant impact on load balances. %

Smart meters form an essential building block in smart grids by providing prosumer data in the form of voltage, current,
and power.
However, 
it is very simple to estimate household occupancy or the number of residents based on this data. With more advanced methods, it has been shown that it is possible to classify electrical devices in use or even which TV channel is watched \cite{asghar2017smart,Molina2010private,Greveler2012multimedia,WOOD2003821,5054916}.
Such information is already a valuable resource that can be exploited for burglary, mass surveillance, or tailored advertisement \cite{jawurek2012privacy}. Furthermore, detailed private information can be disclosed when additional background knowledge is available \cite{Molina2010private}. Thus, leakage of smart meter data poses a major threat. 
At the same time, private companies (with or without regulatory oversight) are partially commissioned to store and process this data. For instance, Itron, Triliant, Landis+Gyr, Oracle, Sensus, and Siemens offer storage and processing solutions for metering data. Besides regulatory measures, technical solutions can help to secure smart metering data against abuse. Hence, institutions such as the European Commission and the US National Institute for Standards and Technology %
stress the importance of introducing privacy-preserving measures to smart grids \cite{EUcommission15} \cite{NISTIR7628}. 

Against this background, we are interested in a privacy-preserving PFA based on confidential smart metering and forecast data. This way, a reliable grid operation and security are
no longer conflicting goals.
Our interest in power flow is additionally driven by its fundamental role in power systems and the fact that solution techniques via Newton's method (which we will use here) are applied across various contexts. Specifically, these methods are pivotal in state estimation, optimal power flow \cite[Chapters 2.6 and 8]{zhu2015optimization}, and advanced grid optimization techniques \cite{faulwasser2018optimal}, underscoring their relevance beyond the scope of this paper.

\subsection{Related work}

Privacy-preservation in smart grids has been addressed in the literature before. We compactly summarize related works and point out conceptual differences to our approach in \Cref{tab:related}.
The listed works can be distinguished based on the applied cryptographic technique,
the application, whether a trusted third party (TTP) is required, and whether some crucial data is leaked.
We also provide a coarse classification of the problem complexity, which reflects the difficulty of implementing the %
task at hand securely, without making use of a TTP or leaking data.
In the following, we discuss some related works~in~more~detail.

Common attempts to alleviate privacy problems are restricting the measurement periods to $15$ minutes or more, limiting the data storage duration, and anonymization \cite{efthymiou2010smart}. However, these methods prevent leveraging all available information, require trust, and provide limited security guarantees. 
Conceivable cryptographic techniques with 
robust security guarantees are differential privacy (DP), homomorphic encryption (HE), and secure multi-party computation (SMPC). 
The main goal of these techniques is to support some sort of function evaluation while maintaining the privacy of the function's inputs.

In DP,
carefully constructed noise
is added to the data, to hide information while some computation is still possible. Consequently, a trade-off between accuracy and security arises.
Applications of differential privacy for smart grids range from meter data aggregation \cite{acs2011have} and clustering, over state estimation \cite{sandberg2015differentially} to recent works on optimal power flow \cite{FiorettoMH20,gough2021preserving}, neural network based voltage forecasting \cite{ToubeauTMKW23}, and load profile synthesis \cite{huang2022dpwgan}.

Unlike DP, SMPC and HE come with stronger cryptographic security guarantees while enabling high accuracy. These properties make them attractive for a scalable solution with better utility.
HE refers to cryptosystems that allow for computation on ciphertexts.
In the context of smart grids, HE
is extensively used for smart meter data aggregation \cite{GarciaJ10,li2010secure}, where \cite{JoKL16} additionally provides
fraud detection. 
Further, economic dispatch \cite{chen2022privacy} and optimal power flow \cite{wu2021privacy} have been addressed. There, the realization builds on intermediate decryption,
which causes information leakages.
Privacy-preserving load forecasting
is considered in \cite{bos2017privacy}.  

The idea of SMPC is to allow joint computation of a function between multiple parties while keeping each party's input secret.
In contrast to HE, the bottleneck in SMPC is usually not computational, but communication complexity. 
For smart grids, \cite{mustafa2019secure} and \cite{kursawe2011privacy} use secret-sharing based aggregation protocols, and
\cite{danezis2013smart} implements a simple statistical analysis for metering data.
An economic dispatch problem is considered in \cite{tian2022fully}, where the optimization is solved partially private using (Shamir's) secret sharing.
Similarly, energy flow is considered in \cite{Si2023ies} in the context of radial alternating current grids, where the authors use a trusted third party to establish consensus between the participants; however, the more complex subproblems involving the power flow equations are solved in plaintext.
Finally, \cite{zhou2023bidirectional,AbidinACM16,Mu2022trading} utilize SMPC to devise privacy-preserving energy markets.

\renewcommand{\theadfont}{\small}%
\begin{table}[h]
    \centering
    \caption{Overview of related work.}
    \label{tab:related}
    \resizebox{\columnwidth}{!}{%
        \begin{tabularx}{1.18\columnwidth}{llXlcc}
        \toprule
        \thead[l]{Work} &
        \thead[l]{Crypt.\\technique} &
        \thead[l]{Application} &
        \thead[l]{Problem \\complexity} &
        \thead[l]{No\\TTP} &
        \thead[l]{No \\leakage} \\
        \midrule
           \cite{efthymiou2010smart} & Escrow & %
           Real-time~metering data\!\!\! & High & \xmark & \xmark \\
           \cite{acs2011have} & DP & Data aggregation & Low & \cmark & \xmark \\
           \cite{sandberg2015differentially} & DP & State estimation & Medium & \cmark & \xmark \\
           \cite{FiorettoMH20} & DP & Power line obfuscation & Medium & \cmark & \xmark \\
           \cite{gough2021preserving} & DP & Optimal power flow & High & \xmark & \xmark \\
           \cite{ToubeauTMKW23} & DP & Load forecast & Medium & \cmark & \xmark \\
           \cite{huang2022dpwgan} & DP & Load profile synthesis & Medium & \cmark & \xmark \\
           \cite{GarciaJ10,li2010secure,JoKL16} & HE & Data aggregation & Low & \cmark & \cmark \\
           \cite{chen2022privacy} & HE & Economic dispatch & Medium & \cmark & \xmark \\
           \cite{wu2021privacy} & HE & Optimal power flow & High & \xmark & \xmark \\
           \cite{bos2017privacy} & HE & Load forecast & Medium & \cmark & \cmark\\ \cite{mustafa2019secure,kursawe2011privacy,danezis2013smart} & SMPC & Data aggregation & Low & \cmark & \cmark \\
            \cite{tian2022fully} & SMPC & Economic dispatch & Medium & \cmark & \cmark \\
            \cite{Si2023ies} & SMPC & Optimal power flow & High & \xmark & \xmark \\
           \cite{zhou2023bidirectional,AbidinACM16} & SMPC & Energy markets & Medium & \cmark & \cmark \\
           \cite{Mu2022trading} & SMPC & Energy markets & Medium & \cmark & \xmark \\
           \midrule
           Ours & SMPC & Power flow analysis & High & \cmark & \cmark \\
         \bottomrule
        \end{tabularx}
    }
\end{table}

\subsection{Contributions}
When comparing the available 
frameworks for
SMPC and HE, it becomes clear that 
those for SMPC are more mature than their counterparts for HE. In addition, due to the heavy computational overhead of HE, SMPC performs much better in settings where the computing parties consist of resource-constrained devices such as smart meters.
Thus, in this work, 
we present the first realization of a fully privacy-preserving PFA based on Newton's method, leveraging
the strong security guarantees of SMPC protocols.
In more detail, the main contributions of our paper can be summarized as follows:

\begin{itemize}
    \item While SMPC allows for arbitrary computations, a careless choice of protocols and parameters or naive implementations of algorithms typically lead to inefficient results. We provide an efficient SMPC implementation tailored for PFA by considering a suitable Cartesian formulation and by exploiting sparsity. Additionally, we adapt solvers for systems of linear equations (such as LU decomposition and GMRES) and integrate line search techniques. We share our implementation for reproducibility. 
    \item %
    We provide a security analysis
    of our algorithm in the universal composability framework (UC) \cite{Canetti01}. This is especially important since PFA is an important subtask in applications such as flexibility markets, demand response, detection of voltage range violations, state estimation, and optimal power flow. UC-security means that our algorithm can be securely used as a component of other secure programs, e.g., for aforementioned complex applications.
    \item We benchmark our algorithm (using both LU and GMRES solvers) with several SMPC protocols for two grids of varied topology and size. In addition, we break out runtime by network latency. Since the utilized SMPC protocols provide differing security guarantees, we can also show the efficiency of our algorithm for various threat models. This enables us to present profound insights into the practicality of SMPC in the context of PFA, and allows us to substantiate the claim that SMPC-based privacy-preserving PFA can be practical for many threat models and real-world settings.
\end{itemize}
Our paper is structured as follows. In Section \ref{sec:prelim}, we briefly recall the power flow problem and introduce the techniques that form the basis of our privacy-preserving implementation. 
Section \ref{sec:tailoredformulation} then deals with a suitable power flow formulation, while Section \ref{sec:computationsandsecurity} and \ref{sec:security} are concerned with its SMPC implementation and security, respectively. Section \ref{sec:numerics} presents the benchmarks for %
two different smart grid topologies. 
We conclude with Section \ref{sec:conclusion}.

\section{Preliminaries}
\label{sec:prelim}

This section serves as a preparation for the remainder of this paper. We will first recall the well-known power flow problem. Subsequently, 
relevant concepts in SMPC are conveyed, and
lastly, our focus shifts towards \textit{secret sharing}, which 
will form the basis for our implementation in Section \ref{sec:computationsandsecurity}.
The observations in \ref{subsec:securempc} provide a basis for the design decisions made in Sections \ref{sec:tailoredformulation} and \ref{sec:computationsandsecurity}.

\subsection{The power flow problem}
Power systems are typically modeled by a set of buses (nodes) and branches (transmission lines). Remarkably, assuming stationary operating conditions in a grid with alternating current, the $N\in\N$ buses are fully described by the voltages $\vb\in\C^{N}$ (typically given in terms of magnitude and phase angle) as well as the active and reactive powers $\pb\in\R^N$ and $\qb\in\R^N$.
In this work, we focus on the privacy of smart grid prosumers. Therefore, only load buses (PQ-buses) are of interest here. In other words, $\vb$ is the unknown quantity while forecasts based on historical data or predictions are available for the active and reactive power vectors $\pb_0$ and $\qb_0$, respectively.

In order to find $\vb$, we first note that it is linearly related to the currents $\ib\in\C^{N}$ by the admittance matrix $\Yb=\Gb+\jc\Bb$ via $\ib=\Yb\vb$.
With this relationship, the bus powers can be equivalently and compactly expressed via the apparent power as
$\sb =\diag(\vb)(\Yb\vb)^\ast$. Finally, with help of the prescribed values $\sb_0=\pb_0 + \jc\qb_0$, we find the relation
\begin{equation}
	\label{eq:powerflow}
	\sb(\vb)-\sb_0=\diag(\vb)(\Yb\vb)^\ast-\pb_0-\jc\qb_0=\zerob,
\end{equation}
which is the well-known power flow
equation.
Note that~\eqref{eq:powerflow} is a non-linear system of equations in $\vb$  for which Newton's method is a popular choice. To this end, \eqref{eq:powerflow} is rephrased in terms of $\Fb(\xb)=\zerob$ (details follow in Section \ref{subsec:cartesianform}).
Then, instead of directly solving $\Fb(\xb)=\zerob$, a sequence of linear systems in the form
\begin{equation}
    \label{eq:newton}
    \Jb(\xb_k)\Delta \xb_k=-\Fb(\xb_k),
\end{equation}
with $k\in\N$ and the Jacobian $\Jb=\partial \Fb/\partial \xb$, is solved until the residual $\|\Fb(\xb_k)\|_2^2$ is sufficiently small.
After solving~\eqref{eq:newton} for $\Delta\xb_k$, the state is updated according to $\xb_{k+1}=\xb_k+\Delta \xb_k$. 

Once voltages $\vb$ satisfying~\eqref{eq:powerflow} are found, the remaining unknown quantities are readily computed, such as
the voltage and branch current magnitudes $\Vb$ and $\Ib_b$. These quantities are of interest for verifying the operating conditions of the grid via
\begin{equation}
    \label{eq:constraints}
    \Vb_{\!\min}\leq \Vb \leq \Vb_{\!\max} \quad \text{and} \quad
    \Ib_{b} \leq \Ib_{\max}.
\end{equation}
In case \eqref{eq:constraints} is violated, the grid's safety and stability is endangered and electrical devices or systems may be damaged. Therefore, it is of paramount importance that the grid operator takes adequate measures to ensure \eqref{eq:constraints} at all times.

\subsection{Overview of secure multi-party computation}
\label{subsec:securempc}

In SMPC, $n$ parties (sometimes also called players) $\P_1,\dots,\P_n$
agree to jointly compute a
function $\yb=\gb(\zb)$, where $\yb = (y_1,\ldots,y_m)^\top$ and $\zb = (z_1,\ldots,z_n)^\top$ with $m,n\in\N$.
Here, $z_i$ denotes $\P_i$'s secret input and $\yb$ a vector of outputs.
The function $\gb$ is said to be computed \textit{securely} if the following two conditions are satisfied:
\begin{itemize}
    \item \textit{Correctness:} The correct result $\yb$ is computed.
    \item \textit{Privacy:} None of the parties obtains any new information about $\zb$, other than what can be inferred from $\yb$.
\end{itemize}

Up to $n-1$ parties may collude to try to learn about the secrets of other parties. We model this by assuming that an adversary that tries to learn about the secret inputs may corrupt up to $t$ of $n$ parties. In the cases $t\leq \floor{n/2}$ and $t\leq n-1$, we say that we assume an \textit{honest majority} or a \textit{dishonest majority} of parties, respectively. Depending on the threat model, corrupted parties may either be
\begin{itemize}
    \item \textit{semi-honest} (also called \textit{honest-but-curious}), meaning they try to learn secret inputs using the function output and the information released to them during the regular execution of the SMPC protocol, but otherwise follow the protocol faithfully, or
    \item \textit{malicious}, meaning they may do all of the above and also deviate arbitrarily in their behavior during protocol execution.
\end{itemize}
Note that the adversary may pool the information gathered by corrupted parties together and (in the malicious model) coordinate the protocol deviations of corrupted parties to learn more information.

In our context, the role of the parties $\P_1,\ldots,\P_n$
can be assigned to the prosumers, while the prosumers' secret power data is collected in $\zb$.
However, revealing the node voltages $\vb$ would clearly breach the prosumers' privacy. Thus, the output $\yb$ must represent a less vulnerable result, for instance checking operating conditions securely via \eqref{eq:constraints}. Then, in case of a violation, the operator receives a stabilization strategy which, of course, must be revealed to it. We omit specifying additional computations and focus on the solution of \eqref{eq:powerflow} here. Thus, $\yb=\gb(\zb)$ can mainly be understood as a mapping of power data to the voltage $\vb$ such that \eqref{eq:powerflow} is satisfied.

\subsection{Instantiation of SMPC}
\label{subsec:instants_SMPC}

\begin{figure}[t]
    \includegraphics[trim=0cm 0cm 0cm 0cm, clip=true, width=\linewidth]{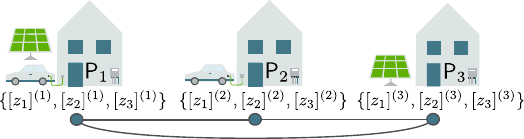}
    \caption{Privacy preserving computation via secret sharing in a smart grid with $n=3$ prosumers. Each prosumer $\Ps_i$ has $3$ shares and is connected via a point to point connection (gray lines) to all other prosumers.} 
    \label{fig:A-SS}
\end{figure}

One approach towards SMPC is secret sharing. It provides a way for a party $\P_i$ to distribute information about a secret $z_i$ across all parties $\P_1,\dots,\P_n$ such that they all together hold full information on $z_i$, yet no party on its own (except $\P_i)$ has any information about $z_i$. 
For simplicity, we consider a three-party scheme at first:
Let $z_1\in\mathbb{Z}_p$ represent $\P_1$'s secret, with $p$ prime.
Note that computations on secret-shared numbers are performed over $\Z_p$, which implies a reductions modulo $p$.
To share the secret, $\mathsf{P}_1$ chooses $r_1,r_2$ uniformly at random in $\mathbb{Z}_p$ and sets $r_3=z_1-r_1-r_2$. Observe that for each $r_1,r_2$ and $r_3$ all values in $\mathbb{Z}_p$ are equally likely, which is why they do not contain any information about $z_1$. Therefore, we can distribute the \textit{shares} $ r_1, r_2$ and $r_3$ to $\mathsf{P}_1,\mathsf{P}_2$ and $\mathsf{P}_3$, respectively, such that no party alone learns something new about $z_1$, but all players together can \textit{reveal} (or ``\textit{open}'') the secret by computing $z_1=r_1+r_2+r_3$. We denote a secret-shared number $z_i$ as $\share{z_i}$ and its $j$'th share as $\share{z_i}^{(j)}$. 
Thus, in the case above, we can also write $\share{z_1}=(\share{z_1}^{(1)},\share{z_1}^{(2)},\share{z_1}^{(3)})=(r_1,r_2,r_3)$. An illustration of the setup is depicted in Figure~\ref{fig:A-SS}.

It is easy to see that $\mathsf{P}_1,\mathsf{P}_2,\mathsf{P}_3$ can compute $z_4=z_1+z_2+z_3$ where $z_1,z_2,z_3$ are arbitrary secret-shared numbers. 
First, each player $\P_i$ (for $i\in\{1,2,3\}$) locally computes $\share{z_4}^{(i)}= \share{z_1}^{(i)}+ \share{z_2}^{(i)}+ \share{z_3}^{(i)}$. Then, to \textit{reveal} $z_4$, they can send each other their shares of $z_4$ and compute $z_4=\share{z_4}^{(1)}+\share{z_4}^{(2)}+\share{z_4}^{(3)}$. From now on we write $\share{z_1}+\share{z_2}$ to mean that each player $\P_i$ computes the sum of their local shares $\share{z_1}^{(i)},\share{z_2}^{(i)}$ as described above. Addition, subtraction and multiplication of a secret-shared number with a public constant, as well as subtraction of two secret-shared numbers, can be computed locally in a similar fashion.

In order to compute the product of two secret-shared numbers, one can use \textit{multiplication triples} \cite{beaver}. Here, we assume that the $n$ players are in possession of secret-shared triples $\share{a},\share{b},\share{c}$ of the form $c = ab$ where $a,b$ are chosen uniformly at random from $\mathbb{Z}_p$. 
Now observe that 
\begin{equation*}
\begin{split}
    z_1 z_2 
    &=(z_1+a-a)(z_2+b-b)\\
    &=(z_1+a)z_2+(z_1+a)b-(z_1+a)b-(z_2+b)a+ab\\
    &=(z_1+a)z_2-(z_2+b)a+c.
    \end{split}
\end{equation*}
Assuming $a,b$ and $c$ are secret, the parties can open $z_1+a$ and $z_2+b$ without any loss of privacy and compute 
\begin{equation*}
    \label{eq:mult_ss}
\share{z_1 z_2}=(z_1+a)\share{z_2}-(z_2+b)\share{a}+\share{c}
\end{equation*}
based on addition and multiplication by a public constant. 
From now on we write $\share{z_1}\share{z_2}$ to mean that $\P_1,\dots,\P_n$ follow a multiplication procedure such as the one described above to securely compute $\share{z_1z_2}$.
 Remarkably, it is possible to express any function in terms of addition and multiplication gates. 
We omit giving details
and refer the interested reader to \cite{catrina2010secure,aly2019benchmarking} for algorithms for standard mathematical functions.

Thus, the approach taken above allows secure computation of arbitrary functions in the \textit{preprocessing model} even with a \textit{dishonest majority} where all but one of the parties are dishonest and colluding. We refer to multiplication triples and other random numbers that are helpful during computation but unrelated to the actual secret inputs or the computed function as \textit{correlated randomness}. 
When we say that a computation is secure in the preprocessing model, we mean that the computation is secure under the assumption that the correlated randomness 
required for the
computation has been generated securely. This is the case if, for example, the correlated randomness has either been generated and distributed by a trusted third party (a so-called \textit{dealer}), or 
if the correlated randomness has been generated interactively among the computing parties using secure protocols based on HE \cite{KellerPR18} or oblivious transfer \cite{KellerOS16}. We refer to the generation of correlated randomness as the \textit{offline} or \textit{preprocessing} phase of the computation, and 
denote 
the input-dependent computation as the \textit{online} phase.
We can avoid the preprocessing model by, for example, using the SMPC protocols BGW/CCD \cite{ben2019completeness}\cite{ChaumCD88}.

Lastly, let us consider the complexity of several operations as a preparation for the following section.
While local computation in secret sharing is typically cheap, communication
constitutes a bottleneck. 
Against this background, the number of \textit{communication rounds} (how often a communication is invoked) is a useful metric.
Recall that additions and multiplications of secret-shared numbers with public constants, and $\share{z_1}+\share{z_2}$
are communication-free. 
On the other hand, $\share{z_1}\share{z_2}$ requires at least one communication round, and (due to truncation \cite{catrina2010secure}) at least two for fixed-point numbers.
Elementary functions that are useful for the solution of the power flow problem \eqref{eq:powerflow} are divisions, trigonometric functions like $\sin(z)$, and square roots of secret-shared numbers.
Based on the methods presented in \cite{aly2019benchmarking}, these require approximately $40$, $40$, and $87$ rounds, respectively, for a $64$-bit number.

\section{Tailored power flow analysis}
\label{sec:tailoredformulation}
Equipped with the knowledge from Section~\ref{sec:prelim}, we can now shift our focus towards a privacy-preserving solution of the power flow problem~\eqref{eq:powerflow} by first identifying a suitable solution algorithm. In fact, one of the main
ingredients for a practical implementation is a solution algorithm that is adapted to the requirements of SMPC protocols.
While the techniques we discuss and employ in this section are known, interdisciplinary knowledge is required to select a well-suited tailored approach.

\subsection{Discussion of different formulations}\label{ssec:formulation}

There are different formulations available for the power flow problem \cite{sereeter2019comparison}. 
First, a complex formulation can be considered, where~\eqref{eq:powerflow} is solved directly with $\vb$ as a variable. However, the derivative of a complex conjugate is not well-defined. Thus, \cite{le1997newton} relies on an approximation based on differentials, which can deteriorate the step direction. In \cite{sereeter2019comparison}, the conjugate currents are kept in the formulation, which leads to linear convergence (opposed to quadratic convergence in other formulations). Thus, although a complex formulation can reduce the solution effort in each iteration, the number of iterations is expected to be higher. 
Second, a polar formulation for \eqref{eq:powerflow} can be used, which is a popular choice. There, $\vb$ is understood in terms of voltage magnitude and phase angle. In this case, the Jacobian will contain sine and cosine terms, which have to be reevaluated
whenever
a phase angle is updated. Especially in an SMPC formulation, this would become a major source of inefficiency.
Although intermediate forms exist, the best suited option seems to be a Cartesian formulation because the Jacobian can be implemented
using only additions and multiplications, which are native SMPC operations.

\subsection{Cartesian power flow formulation}
\label{subsec:cartesianform}
Since a Cartesian formulation is well-known we will only briefly specify the remaining quantities in \eqref{eq:newton}. First, real and complex components are
treated separately, which leads to the definitions 
\begin{align*}
    \xb = 
    \begin{pmatrix}
       \vb_R \\ \vb_I
    \end{pmatrix}
    = 
    \begin{pmatrix}
        \Re\vb \\
        \Im\vb
    \end{pmatrix}
    \quad \text{and} \quad
    \Fb(\xb) = \begin{pmatrix}
    \pb(\xb)-\pb_0 \\
    \qb(\xb)-\qb_0
    \end{pmatrix}.
\end{align*}
Next, we proceed 
with $\sb=\diag(\vb)(\Yb\vb)^\ast$ and obtain
\begin{equation}
\label{eq:powerandvoltage}
    \begin{pmatrix}
        \pb(\xb) \\
        \qb(\xb)
    \end{pmatrix}
    \!=\!
    \begin{pmatrix}
        \diag(\vb_R) & \!\!\! \blind{-}\diag(\vb_I) \\
        \diag(\vb_I) & \!\!\! -\diag(\vb_R)
    \end{pmatrix}
    \!\!
    \begin{pmatrix}
        \Gb & \!\! -\Bb \\
        \Bb & \!\! \blind{-}\Gb
    \end{pmatrix}
    \!\!
    \begin{pmatrix}
        \vb_R \\
        \vb_I
    \end{pmatrix}
\end{equation}
which specifies $\Fb(\xb)$ and 
\begin{align}
\label{eq:jacobian}
    &\Jb(\xb) = \begin{pmatrix}
        \Hb_1 + \Hb_3 & \blind{-}\Hb_2 + \Hb_4 \\
        \Hb_2 - \Hb_4 & -\Hb_1 + \Hb_3
    \end{pmatrix}
    \quad \text{with} \\
    \nonumber
    \Hb_1\!= &\diag(\vb_R)\Gb + \diag(\vb_I)\Bb, \;\, \Hb_3\!= \diag(\Gb\vb_R -\Bb\vb_I), \\
    \nonumber
    \Hb_2\!= &\diag(\vb_I)\Gb -\diag(\vb_R)\Bb,  \;\, \Hb_4\!= \diag(\Bb\vb_R +\Gb\vb_I).
\end{align}
In order to avoid ambiguous solutions for $\Fb(\xb)=\zerob$, the voltage at the so-called slack node must be fixed, which serves as a voltage reference. Here, the slack node is, without loss of generality, assumed to be the first node in the grid, and the corresponding rows in $\Fb$ as well as the rows and columns in $\Jb$ are omitted.

\subsection{Line search}
\label{subsec:linesearch}
Instead of using the solution of~\eqref{eq:newton} directly as a Newton step, we introduce the step size $\eta$ such that the update becomes $\xb_{k+1}=\xb_{k}+\eta \Delta \xb_{k}$. By choosing $\eta$ suitably, the convergence speed and the robustness of Newton's method are enhanced. %
Finding an appropriate $\eta$ is usually done by approximating the optimal step size
\begin{equation}
    \label{eq:linesearch}
    \eta^{\star}
    =\argmin_{\eta} g(\eta)
    =\argmin_{\eta} \frac{1}{2} \Fb(\eta)^\top \Fb(\eta),
\end{equation}
where we omitted the dependency on $\xb_{k}$ as well as $\Delta \xb_{k}$, since they are constant in~\eqref{eq:linesearch}. 
Of course, algorithms for \eqref{eq:linesearch} exist (see for instance~\cite[Chapter 3]{nocedal1999numerical} and~\cite[Chapter 9]{press2007numerical}). However, they require additional elementary functions and logic, which increases the communication overhead in an SMPC solution. Instead, we opted for the following approach.
\\
The optimizer of \eqref{eq:linesearch} satisfies $\mathrm{d}g(\eta^\star)/\mathrm{d}\eta = 0$ where a change by $\Delta \eta_j = \eta_{j+1}-\eta_{j}$ with $j\in\N$ is
\begin{equation}
    \label{eq:firstorderapprox}
    \frac{\mathrm{d} g(\eta_j + \Delta \eta_j)}{\mathrm{d}\eta} \approx \frac{\mathrm{d} g(\eta_j)}{\mathrm{d}\eta} +\frac{\mathrm{d}^2 g(\eta_j)}{\mathrm{d}\eta^2}  \Delta \eta_j  = 0.
\end{equation}
Then, the finite difference formulas
\begin{align*}
\frac{\mathrm{d}g(\eta_j)}{\mathrm{d}\eta} = 
\Fb(\eta_j)^\top &  \Jb(\eta_j) \Delta \xb_k \approx \frac{1}{\eta_j} \Fb(\eta_j)^\top \left(\Fb(\eta_j)-\Fb(0)\right),
\\
\frac{\mathrm{d}^2 g(\eta_j)}{\mathrm{d}\eta^2} &\approx \frac{\mathrm{d}g(\eta_j)/\mathrm{d}\eta-\mathrm{d}g(\eta_{j-1})/\mathrm{d}\eta}{\eta_{j}-\eta_{j-1}},
\end{align*}
and the abbreviation $\xi_j :=\Fb(\eta_{j})^\top \left(\Fb(\eta_{j})-\Fb(0)\right)$ allow rearranging \eqref{eq:firstorderapprox} into the compact form
\begin{align}
\label{eq:step}
    \Delta \eta_{j}
    =
    \Delta \eta_{j-1} \frac{\eta_{j-1} \xi_j}{\eta_j \xi_{j-1}-\xi_j}.
\end{align}
Note that in \eqref{eq:step} computing $\Jb$ is circumvented and allowing $\Delta\eta>1$ can further reduce the amount of linear equation systems~\eqref{eq:newton} to be solved. 

\section{Privacy-preserving power flow analysis}
\label{sec:computationsandsecurity}

The following section deals with a privacy-preserving solution of~\eqref{eq:powerflow} using the formulation and tools discussed in Sections \ref{subsec:instants_SMPC} and  \ref{sec:tailoredformulation}, respectively. 
\\
In order to simplify the presentation we use $[x]$ on $x\in\R$ to denote the fixed-point encoding in $\Z_p$ via $z = \roundtext{2^h x}\bmod p$ for some $h\in\N$ and a subsequent secret sharing.
Initially, each prosumer $\Ps_i$ computes shares $(\share{p_{0,i}},\share{q_{0,i}})$ and distributes them among the other $n-1$ parties.
Similarly, shares of the initial guess $\share{\xb_0}:= ( \share{\vs_{R,1}} \dots, \share{\vs_{R,N}},\share{\vs_{I,1}},\dots \share{\vs_{I,N}})^\top$
and (depending on their secrecy) shares or the public values of $\Gb,\Bb$ are distributed. 
In case $\Gb$ and $\Bb$ are unavailable at first, one may use the estimation strategy proposed in Appendix \ref{app:admittance}.

\subsection{Gradient vector and Jacobian}
Next, $\share{\Fb(\xb)}$ and $\share{\Jb(\xb)}$ are computed where we drop the iteration index $k$ and consider $i\in\{1,2,\dots,N\}$ (for the moment) to simplify the presentation.
Given $\share{\xb}$, the prosumers first invoke addition and multiplication protocols to compute
\begin{align}
\label{eq:sharecurrents}
\begin{split}
        \share{i_{R,i}} &= \sum_{j=1}^{N} \share{G_{ij}} \share{\vs_{R,j}} - \share{B_{ij}} \share{\vs_{I,j}}  \\
        \share{i_{I,i}} &= \sum_{j=1}^{N} \share{B_{ij}} \share{\vs_{R,j}} + \share{G_{ij}} \share{\vs_{I,j}}.
\end{split}
\end{align}
Then, with $\share{i_{R,i}}, \share{i_{I,i}}$ at hand, one can compute $\share{\Fb(\xb)} := (\share{F_1},\dots,\share{F_{2N}})^\top$ via
\begin{align}
\label{eq:shareF}
\begin{split}
 \share{F_i} &= \share{\vs_{R,i}} \share{i_{R,i}} + \share{\vs_{I,i}} \share{i_{I,i}} - \share{p_{0,i}} \\ 
 \share{F_j} &= \share{\vs_{I,i}} \share{i_{R,i}} - \share{\vs_{R,i}} \share{i_{I,i}} - \share{q_{0,i}}
\end{split}
\end{align}
where $j=i+N$.
Lastly, we focus on the shares in $\share{\Jb(\xb)}$ which depend on
\begin{align}
\label{eq:shareblockH}
\begin{split}
    \share{H_{1,ij}} &= \share{G_{ij}} \share{\vs_{R,i}} +  \share{B_{ij}}\share{\vs_{I,i}} \\
    \share{H_{2,ij}} &=  \share{G_{ij}} \share{\vs_{I,i}}-  \share{B_{ij}}\share{\vs_{R,i}} \\
    \share{H_{3,ii}} &=\share{i_{R,i}} \quad \text{and} \quad H_{3,ij} = 0 \quad \text{for $i\neq j$}\\
    \share{H_{4,ii}} &= \share{i_{I,i}}\, \quad \text{and} \quad H_{4,ij} = 0 \quad \text{for $i\neq j$}
\end{split}
\end{align}
where $\share{G_{ii}} \share{\vs_{R,i}}$, $\share{G_{ii}}\share{\vs_{I,i}}$, $\share{B_{ii}}\share{\vs_{R,i}}$, $\share{B_{ii}}\share{\vs_{I,i}}$ can be reused from \eqref{eq:sharecurrents}. %
Finally, we collect the shares and construct $\share{\Jb(\xb)}$ as in \eqref{eq:jacobian} via the local addition protocol.
\\
\begin{complexity}
The execution of \eqref{eq:sharecurrents} requires at most $4N^2-4N$ secret multiplications. However, they can be reduced as follows. First, the sparsity structure of $\Gb$ and $\Bb$ may be public information (because it is directly linked to the grid's topology). Then, only multiplications with non-zero factors have to be performed. 
Second, by assuming $\Gb$ and $\Bb$ are public, all multiplications in \eqref{eq:sharecurrents} become local operations. Although this assumption is not strictly necessary, it is reasonable from our perspective since $\Gb$ and $\Bb$ do not contain confidential information (they could be calculated by the prosumers with little effort). 
\\
Considering \eqref{eq:shareF} and \eqref{eq:shareblockH} note that $F_i$ and $J_{ij}$ for $i,j\in\{1,N+1\}$ can be omitted due to the slack node.
Then, \eqref{eq:shareF} for $\share{\Fb(\xb)}$ costs $4N-4$ multiplications.
Similarly, \eqref{eq:shareblockH} for $\share{\Jb(\xb)}$ depends mostly on \eqref{eq:sharecurrents}. Taking the precomputed results into account, the Jacobian requires at most $4(N-1)^2-4N$ additional multiplications.
However, for the computation of $\share{H_{1,ij}},\share{H_{2,ij}}$ also potential savings due to public $\Gb,\Bb$ apply.
Interestingly, if $\Gb = \Gb^\top$ and $\Bb = \Bb^\top$ then the previous results from \eqref{eq:sharecurrents} determine \eqref{eq:shareblockH} even if $\Gb$ and $\Bb$ are secret, i.e., \eqref{eq:shareblockH} becomes free.
\\
In total, the computation of $\share{\Fb(\xb)}$ and $\share{\Jb(\xb)}$ requires at most $8N^2-12N$ multiplications. 
The fastest case arises if $\Gb,\Bb$ are not kept secret, in which \eqref{eq:sharecurrents} and \eqref{eq:shareblockH} become local operations and can be computed with $4N-4$ multiplications.
\end{complexity}

\subsection{Solving the system of linear equations}
Here, we consider solving $\Jb(\xb)\Delta \xb = - \Fb(\xb)$ but using  $\share{\Jb(\xb)}$ and $\share{\Fb(\xb)}$.
Taking into account that $\Jb$ offers no special properties (other than sparsity because a bus in low voltage grids has typically not more than two neighbors), we will experimentally compare a suitable direct with an indirect solver. Namely, an LU decomposition \cite[Chapter 2.3]{press2007numerical} with a subsequent forward-backward substitution, and GMRES~\cite[Algorithm~6.9]{saad2003iterative}. Both have been slightly optimized for an SMPC execution.
\\
\subsubsection{Direct solver}
Algorithm \ref{alg:directsolve} illustrates our LU decomposition and the forward-backward substitution based on shares. As a reminder, the LU decomposition computes $\share{\Lb},\share{\Ub}$ such that $\Lb \Ub = \Jb(\xb)$. This allows to solve $\Lb \Delta \tilde{\xb} = - \Fb(\xb)$ first and then $\Ub \Delta \xb = \Delta \tilde{\xb}$ where the triangular structure of $\Lb$ and $\Ub$ is beneficial. 
Note that we refrain from pivoting (swapping rows and/or columns) to save communications (mainly for comparisons). While no pivoting is generally not recommendable due to numerical instability, our strategy can be justified a priori as shown in Appendix \ref{app:pivot}.
\\
\begin{algorithm}[h]
\caption{Direct solver}
\textbf{Input:} $\share{\Jb(\xb)}$ where $\Jb\in\R^{l\times l}$. \textbf{Outputs:} $\share{\Lb},\share{\Ub}$ \\
Initialize $\share{\Ub} = (\share{\Jb_{1,:}} \;\;\share{\zerob_{l \times l-1}})$ and 
\\
$\share{\Lb} = (\share{\Jb_{:,1}}/\share{J_{1,1}} \;\; ( \share{\zerob_{l-1\times 1}} \;\; \share{\mathbb{I}_{l-1\times l-1}} )^\top )$  %
\begin{algorithmic}[1]
        \For{$i = 2$ to $l$}  \Comment{LU decomposition}
        \For{$j = i$ to $l$}
            \State{$\share{U_{ij}} = \share{J_{ij}} - \sum_{k=1}^{i-1} \share{L_{ik}} \share{U_{kj}}$}
        \EndFor
        \For{$j = i+1$ to $l$}
                \State{$\share{L_{ji}} = (\share{J_{ji}} - \sum_{k=1}^{i-1} \share{L_{jk}} \share{U_{ki}})/\share{U_{ii}}$}
        \EndFor
        \EndFor
\For{$i = 1$ to $l$} \Comment{forward-backward substitution}
   \State{$\share{\Delta \tilde{x}_i} = \share{F_i} - \sum_{j=1}^{i-1} \share{L_{ij}} \share{\Delta \tilde{x}_j}$} %
\EndFor
\For{$i=l$ to $1$}
    \State{$\share{\Delta x_i} = (\share{\Delta \tilde{x}_i} - \sum_{j= i+1}^l \share{U_{ij}} \share{\Delta x_j} )/\share{U_{ii}}$}
\EndFor
\end{algorithmic}
\label{alg:directsolve}
\end{algorithm}
\\
\begin{complexity}
    By omitting the slack node, we have $l=2N-2$.
    Here, we picked the $l$ degrees of freedom offered by the LU decomposition to select $L_{ii}=1$, which spares us $l$ divisions (that would occur in line 10) and $(l^2-l)/2$ multiplications (due to $i+1$ in line 5). The initialization allows starting the iteration at $i=2$ in line 1.
    Additionally, we separated the for-loops for $\share{U_{ij}}$ and $\share{L_{ji}}$ for reasons that become clear in Section \ref{subsec:computation}.
    Now, line 10 and 11 require each $(l^2-l)/2$ sequential multiplications while $1/\share{U_{ii}}$ can be reused. 
    \\
    Thus, Algorithm \ref{alg:directsolve} requires $l$ multiplicative inverses and at most $l^3/3 -l^2/2 + l/6$ multiplications. 
    Further optimizations are possible if the sparsity structure of $\Jb$ is public. This enables to calculate which $L_{ij},U_{ij}$ will be zero beforehand such that corresponding multiplications can be omitted.
    Finally, it can sometimes be effective to reuse $\Lb\Ub=\Jb(\xb_{k-1})$ as a factorization for $\Jb(\xb_{k})$. Unfortunately, this turned out to be not expedient in our numerical experiments.
\end{complexity}    
\\
\subsubsection{Indirect solver} %
In comparison to a direct solver, an iterative approach is utilized by indirect solvers that refine the solution. Here, we opted for GMRES which requires additions, multiplications, divisions, and square roots.
However, we omit to present the pseudocode of our GMRES implementation here because the algorithm is significantly more tedious while an analysis similar to before can be applied. Nonetheless, the interested reader can find details in our code\textsuperscript{\ref{fn:code}}.
\\
Importantly, the convergence speed of GMRES crucially depends on the condition of $\Jb$. Thus, a preconditioner $\Pb$ is typically used such that $\Pb\Jb \Delta\xb=-\Pb\Fb$ is solved instead of~\eqref{eq:newton}. 
However, if $\Pb$ is not specifically constructed, then $\Pb\Jb$ loses a lot of $\Jb$'s sparsity. In this context, we investigated different preconditioners during our numerical experiments in Section \ref{sec:numerics}.
Surprisingly, precomputing $\Pb=\Jb(\xb_0)^{-1}$, where $\xb_0$ is selected close to the nominal operating conditions of the grid, and simply reusing $\Pb$ throughout all iterations led to the best results. Furthermore, warm-starting the GMRES with $\Delta \xb_{k-1}$ in the $k$-th step was not advantageous in our experiments.

\subsection{Line search}
Lastly, we focus on the SMPC realization of the line search procedure from Section \ref{subsec:linesearch}, which is presented in Algorithm \ref{alg:linesearch}.
\begin{algorithm}[h]
\caption{Line search}
\textbf{Inputs:} $\share{\xb},\share{\Delta\xb},\share{\Fb(0)}$ where $\xb,\Fb\in\R^{l}$. \textbf{Output:} $\share{\xb_{k+1}^\star}$ \\
Initialize $j=1$, $\share{\eta_{1}} = \share{0.95}$, $\share{\eta_{0}} = \share{1}$ %
\begin{algorithmic}[1]
\State{$\share{x_i^\prime} = \share{x_{i}} + \share{\eta_{j-1}} \share{\Delta x_{i}}$ \text{for all $i$}}
         \State{$\share{\Fb(\eta_{j-1})}$ by executing \eqref{eq:sharecurrents} and \eqref{eq:shareF}}
         \State{$\share{\xi_{j-1}} =  \sum_{i=1}^{l} \share{F_i(\eta_{j-1})} \left(\share{F_i(\eta_{j-1})}-\share{F_i(0)}\right)$}
\While{$j \leq \ell_{\mathrm{ls}}$}
    \State{$\share{x_{i}^\prime}  = \share{x_i} +\share{\eta_{j}} \share{\Delta x_i}$ for all $i$} 
    \State{$\share{\Fb(\eta_{j})}$ by executing \eqref{eq:sharecurrents} and \eqref{eq:shareF}}
    \State{$\share{\xi_j} = \sum_{i=1}^{l} \share{F_i(\eta_j)}\left(\share{F_i(\eta_j)}-\share{F_i(0)}\right)$} 
    \State{$\share{\Delta \eta_j} = \share{\Delta \eta_{j-1}} \share{\eta_{j-1}}\share{\xi_j}/(\share{\eta_j}\share{\xi_{j-1}}-\share{\xi_j}) $}
    \State{$\share{\eta_{j+1}} = \share{\eta_j} + \share{\Delta \eta_j} $} 
    \State{$j \leftarrow j+1$}
\EndWhile
\State{$\share{x_{k+1,i}} = \share{x_{i}} + \share{\eta_{j+1}} \share{\Delta x_{i}}$ for all $i$}
\end{algorithmic}
\label{alg:linesearch}
\end{algorithm}
Here, $\ell_{\mathrm{ls}}$ is a predetermined number of iterations.
Note that large $\Fb(\xb)$ can lead to numerical issues in line 7 due to the fixed-point encoding. This issue is, however, easily addressed by replacing $\Fb(\xb)$ with $\epsilon \Fb(\xb)$, where $\epsilon$ is small and positive, which has no effect on line 8.

\begin{complexity}
    Again, by omitting the slack node, we have $l=2N-2$.
    Observe that $\Fb(0)$ and for $j>1$ also old $\Fb(\eta_{j})$ can be reused. Then, the multiplication cost of one iteration is given by the cost of \eqref{eq:sharecurrents}, \eqref{eq:shareF} (detailed above) and $2l+4$ (again with $l=2N-2$) multiplications.
    Additionally, one division in line 8 is needed.
\end{complexity}

\subsection{Overview and communication}
\label{ss:privacy_overview}
\label{subsec:computation}
Finally, our privacy-preserving PFA denoted by $\Pi_{\mathsf{PFA}}$ is recapitulated in Algorithm \ref{alg:powerflow} where $\ell$ is a predetermined bound on the number of Newton iterations required for convergence. This prevents correlating the number of iterations with the residuum $\|\Fb(\xb_k)\|_2^2$.
\begin{algorithm}[h]
\caption{Power flow analysis $\Pi_{\mathsf{PFA}}$}
\textbf{Inputs:} $\share{\xb_0}, \share{\pb_0}, \share{\qb_0}, \share{\Gb}, \share{\Bb}.$ \textbf{Output:} $\share{\xb_{k+1}}$ \\
Initialize $k=1$
\begin{algorithmic}[1]
        \While{$k \leq \ell$}
        \State{$\share{\Jb(\xb_k)}$, $\share{\Fb(\xb_k)}$ by  \eqref{eq:sharecurrents},\eqref{eq:shareF}, \eqref{eq:shareblockH}}
        \State{ $\share{\Delta \xb_k}$ via Algorithm \ref{alg:directsolve} or GMRES}
        \State{ $\share{\xb_{k+1}}$ by Algorithm \ref{alg:linesearch}}
        \State{ $k \leftarrow k+1$ }
        \EndWhile
\end{algorithmic}
\label{alg:powerflow}
\end{algorithm}

Since a communication round takes at least \qty{1}{\milli\second} under realistic conditions, just naively computing operations sequentially will lead to inefficient implementations with great communication complexity.
Thus, it is imperative to batch communication rounds as much as possible. Concretely, that means that we will delay communication until a point is reached where all local operations that do not depend on communication have been processed. 
Considering the aforementioned algorithms, the multiplications for \eqref{eq:sharecurrents} can be batched entirely. Then, $\share{\Fb(\xb)}$ and $\share{\Jb(\xb)}$ can be batched in two rounds after the execution of \eqref{eq:sharecurrents} which reduces to one round for public $\Gb,\Bb$.
Furthermore, the multiplications in lines 3 and 6 of Algorithm \ref{alg:directsolve} and the multiplications in Algorithm \ref{alg:linesearch} can be batched into one and three rounds, respectively.

In addition, we need to optimize the latency of our connections. In our observation, the runtime of secure PFA grows in proportion to the RTT of the connection between the participating parties. Thus, prosumers should be connected to each other via a LAN which makes RTTs of around \qty{1}{\milli\second} possible.

\section{Security}
\label{sec:security}

Below, we describe threat models and provide a security analysis in the Universal Composability framework.

\subsection{Threat model and assumptions}
Since our algorithm can be used with various SMPC protocols that provide differing security guarantees concerning the number of corrupted parties and adversarial behavior (as explained in \Cref{subsec:securempc}), our algorithm can be flexibly employed for virtually all threat models. In \Cref{ssec:benchmarks} we provide benchmarks for example grids segmented by security guarantees. 

We assume that the computation is run between all prosumers $\P_1,\dots,\P_n$ in a given smart grid, and the inputs $p_{0,i},q_{0,i}$ of each party $\P_i$ are considered secret. 
The grid operator is only involved in the computation insofar as it may provide $\Gb$ and $\Bb$. Moreover, we assume that the prosumers communicate exclusively via end-to-end encrypted and point-to-point authenticated channels during the computation.
Some SMPC protocols require correlated randomness, which can be provided by a trusted third party, or be generated interactively between the computing parties. 
Again, revealing the output of the PFA will also reveal the secret inputs. Hence, the output should %
be
treated as an intermediary result to realize other privacy-preserving applications, e.g., in the area of flexibility markets, demand response, optimal power flow, or grid management. 

\subsection{Security analysis}

Since we use SMPC protocols that are already proven secure and are able to compute arbitrary functions, we
inherit the security of these SMPC protocols. 
We refer to \Cref{tab:smpc_protocols} to give an overview of the protocols used, their concrete security guarantees, and references to their security proofs.
\begin{table}[h]
{
\caption{Security guarantees of the SMPC protocols used by us.}
\label{tab:smpc_protocols}
\begin{tabularx}{\linewidth}{@{} lllll @{}} 
\toprule
Protocol
& Majority
& Adversary
& {Security Type}
& {Security Proof}\!\!\\
\midrule
Shamir & Honest & Semi-honest\!\! & Perfect & \cite[Thm. 4.2]{AsharovL17}\!\! \\
Atlas & Honest & Semi-honest\!\! & Perfect & \cite[Lem. 1]{GoyalLOPS21} \\
Semi & Dishonest\!\! & Semi-honest\!\! & Computational & \cite[Thm. 5]{KellerOS16} \\
Temi & Dishonest\!\! & Semi-honest\!\! & Computational & \cite[Thm. 3]{Cramer2001multiparty} \\
Sy-Shamir\!\!  & Honest & Malicious & Statistical & \cite[Thm. 4.3]{ChidaHIKGLN23}\!\!\\
Mascot & Dishonest\!\! & Malicious & Computational & \cite[Thm. 5]{KellerOS16}\\
SPD$\Z_{2^k}$ & Dishonest\!\! & Malicious & Computational & \cite[Thm. 2]{CramerDESX18}\\
\bottomrule
\end{tabularx}}
\end{table}
All security proofs are in the \textit{Universal Composability} (UC) framework \cite{Canetti01}. In the following, we will explain how it applies to \Cref{alg:powerflow}. 
UC makes a distinction between the real world where a protocol $\Pi$ is executed between the parties $\P_1,\dots,\P_n$ (of which a subset is controlled by the adversary), and an ideal world, where a trusted party provides $\Pi$\xspace's \textit{functionality} \ucf on the inputs of $\P_1,\dots,\P_n$. $\Pi$ is considered secure, if an environment \ucz controlling the adversary, and providing the inputs to the computing parties, cannot distinguish whether it is interacting with the real world, or with the ideal world where a simulator \ucs with access to \ucf emulates an execution of $\Pi$. Formally:
\begin{defn}[UC-security, Def.~4.19 in \cite{Cramer2015secure}]
    We say that a protocol $\Pi$ securely realizes a functionality \ucf if there exists an efficient simulator \ucs such that:
    \begin{description}%
        \item[Perfect security:] No environment \ucz can distinguish whether it is interacting with $\Pi$ in the real world, or with \ucs in the ideal world.
        \item[Statistical security:] No environment \ucz can distinguish whether it is interacting with $\Pi$ in the real world, or with \ucs in the ideal world, except with probability negligible in $\secpar_1$.
        \item[Computational security:] No efficient environment \ucz can distinguish whether it is interacting with $\Pi$ in the real world, or with \ucs in the ideal world, except with probability negligible in $\secpar_2$.
    \end{description}
\end{defn}
Here, we consider a function $\negl$ to be \textit{negligible}, if for every positive polynomial function $\poly$, there exists an integer $n_0$ such that, when $\secpar > n_0$, we will have 
$
	\negl < \frac{1}{\poly}.
$
Further, we denote an algorithm as \textit{efficient} if it has \textit{probabilistic-polynomial} runtime.

The advantage of protocols shown to be secure in UC is that they remain secure when combined with other UC-secure protocols, and under repeated interaction with multiple users. This is provided by the following theorem.
\begin{thm}[Composition Theorem, Thm.~4.20 in \cite{Cramer2015secure}]
\label{thm:uc}
    Let the composition of $\Pi_\ucf$ and \ucr be a protocol instantiating a functionality \ucf with perfect/statistical/computational security, and let $\Pi_\ucr$ be a protocol instantiating the functionality \ucr with the same type of security. Then the composition of $\Pi_\ucf$ and $\Pi_\ucr$ securely realizes the functionality \ucf with the same type of security.
\end{thm}
All SMPC protocols we use to benchmark our privacy-preserving PFA provide a set of UC-secure arithmetic operations (called the \textit{arithmetic black box} (ABB)), originally introduced in \cite{Cramer2001multiparty}), that comprises secret-sharing a number, addition, multiplication, and revealing of secret-shared numbers, as follows. 
Its ideal functionality $\ucf_{\mathsf{ABB}}$ is defined as follows. 
\begin{defn}[$\ucf_{\mathsf{ABB}}$, {\cite[Figure 7]{KellerOS16}}]\
    \begin{description}
        \item[Input:] On input ($\mathsf{Input},\P_i,\mathsf{id},x$) from $\P_i$ and ($\mathsf{Input},\P_i,\mathsf{id}$) from all other parties, with $\mathsf{id}$ a fresh identifier and $x\in\Z_p$, store ($\mathsf{id},x$).
        \item[Add:] On command ($\mathsf{Add},\mathsf{id}_1,\mathsf{id}_2,\mathsf{id}_3$) from all parties (where $\mathsf{id}_1,\mathsf{id}_2$ are present in memory), retrieve ($\mathsf{id}_1,x$), ($\mathsf{id}_2,y$) and store ($\mathsf{id}_3,x+y$).
        \item[Multiply:] On command ($\mathsf{Mult},\mathsf{id}_1,\mathsf{id}_2,\mathsf{id}_3$) from all parties (where $\mathsf{id}_1,\mathsf{id}_2$ are present in memory), retrieve ($\mathsf{id}_1,x$), ($\mathsf{id}_2,y$) and store ($\mathsf{id}_3,x\cdot y$). 
        \item[Output:] On input ($\mathsf{Output},\mathsf{id}$) from all honest parties (where $\mathsf{id}$ is present in memory), retrieve ($\mathsf{id},y$) and output it to the adversary.
    \end{description}
\end{defn}
Note that the $\mathsf{ids}$ in above definition are an artifact of the UC security proof.
Since even complex operations such as comparisons, 
divisions, and square roots can be derived from the basic operations in the ABB \cite{aly2019benchmarking}, %
it is sufficient for implementing 
\Cref{alg:powerflow}. 
Hence, our privacy-preserving PFA inherits the UC-security of the SMPC protocols we are using. 
\begin{defn}[Ideal functionality of power flow analysis $\ucf_{\mathsf{PFA}}$]
\label{defn:pfa}
    Given a set of inputs $p_{0,i},q_{0,i}$ by each party $\P_i$, the ideal functionality $\ucf_{\mathsf{PFA}}$ outputs the state of the grid in terms of its voltages $\vb$.
\end{defn}
Note that, as pointed out above, the output of $\ucf_{\mathsf{PFA}}$ should not be revealed, but rather be used by another secure algorithm as an intermediary result.
\begin{thm}
\label{thm:pfa}
Let $\Pi_{\mathsf{PFA}}$ be the privacy-preserving PFA defined in \Cref{alg:powerflow}. Then $\Pi_{\mathsf{PFA}}$ realizes $\ucf_{\mathsf{PFA}}$ with 
$\{$perfect/statistical/computational\} security against a \{semi-honest/malicious\} adversary and a \{honest/dishonest\} majority of parties 
depending on the employed SMPC protocol as specified in \Cref{tab:smpc_protocols}.
\end{thm}
\Cref{thm:pfa} follows immediately from \Cref{thm:uc} since each operation from the ABB is UC-secure, and $\Pi_{\mathsf{PFA}}$ only uses operations from the ABB.

\section{Numerical results}
\label{sec:numerics}

Below, we present our secure implementation and provide runtime benchmarks.

\subsection{Example smart grids}

\begin{figure}[tp!]
    \centering
    \includegraphics[trim=4.7cm 10cm 3.9cm 10cm, clip=true,width=.96\linewidth]{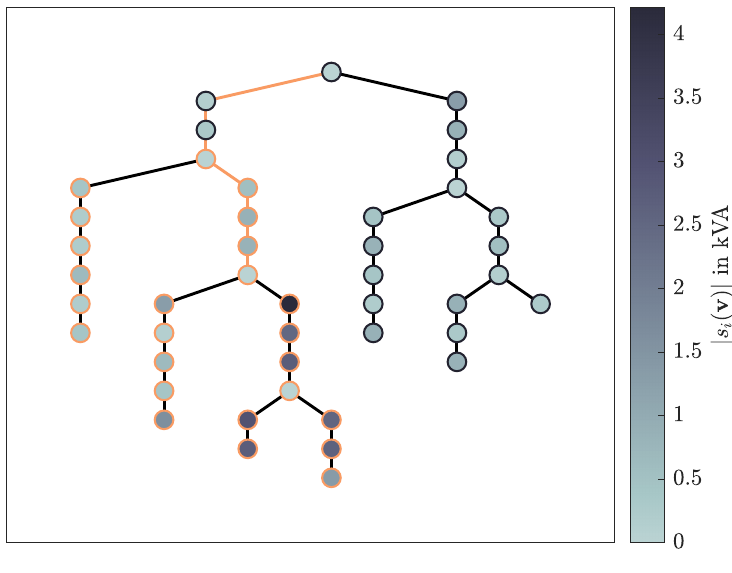}
    \caption{Semi-urban network topology with 44 nodes, of which 39 represent prosumers. The face color of the $i$-th node corresponds to the magnitude $|s_i(\vb)|$, whereas orange node outlines and edges indicate voltage and current constraint violations \eqref{eq:constraints}, respectively.}
    \label{fig:topology}
\end{figure}

For comparability, we selected two low-voltage grids provided by \cite{meinecke2020simbench}\footnote{Datasets \texttt{1\_LV\_semiurb4\_0\_no\_sw} and \texttt{1\_LV\_rur\-al\-1\-\_\-0\_\-no\-\_sw} at time steps $2740$ and $13871$ from \href{https://simbench.de/de/datensaetze/}{https://simbench.de/de/datensaetze/}.}, namely a semi-urban and a rural network which consist of $44$ and $18$ buses with $39$ and $13$ prosumers, respectively.
The former is depicted in Figure~\ref{fig:topology}. We deliberately investigated the worst-case deviation from desired operating conditions. Consequently, the complexity of our computation may be viewed as an upper bound for PFA in the particular grids.
For both grids, we initialize the solution of \eqref{eq:powerflow} via Newton iterations \eqref{eq:newton} with $\xb_0=(\vb_R^\top, \vb_I^\top)^\top=(230,\ldots,230,0, \ldots,0)^\top $.
The step size $\eta_j$ is initialized via $\eta_0=1$ and $\eta_1=0.99$. During the computation, it is typically larger than $1$ for the first iterations and then decays to $0$. Compared to constant step sizes, our line search strategy reduced the required Newton iterations \eqref{eq:newton} by $20-30\%$ in the experiments. 
As already noted, the grids mainly differ in their size.
The Jacobian $\Jb$ of the semi-urban network has the size $86\times 86$, with $7\%$ non-zero elements. 
The worst-case condition number of $\Jb(\xb_k)$ is approximately $8000$, while $\Pb(\xb_0) \Jb(\xb_k)$ has $40\%$ non-zero elements and a condition of at most $30$. 
Meanwhile, for the rural network, one finds that the Jacobian has the size $34 \times 34$, $17\%$ non-zero elements, and a worst-case condition number of (again) $8000$ while $\Pb(\xb_0) \Jb(\xb_k)$ has a density of $90\%$ and a condition of at most $12$.

\subsection{Implementation details}

The code for our algorithms, including compilation instructions, can be found on GitHub\footnote{\href{https://github.com/jvdheyden/privacy-preserving-power-flow}{https://github.com/jvdheyden/privacy-preserving-power-flow}\label{fn:code}}. We make use of the framework \textit{Multi-Protocol SPDZ} (MP-SPDZ)~\cite{keller2020mp} which allows benchmarking secure programs using a variety of SMPC protocols based on secret sharing.
MP-SPDZ also takes care of switching from arithmetic to binary representation when it is advantageous \cite{0001GKRS20} and represents real numbers (as used in PFA) as members of a 
finite set (e.g., $\Z_p$) over which most secret sharing schemes are defined. Furthermore, it helps to batch the communication to reduce the total number of communication rounds.
We set our parameters for computational and statistical security to commonly accepted standard values \cite{aly2019benchmarking,Escudero22}: If the security of an SMPC protocol relies on computational assumptions, our computational security parameter is $\secpar_2=128$, meaning that an efficient adversary has a success probability smaller than $2^{-128}$. Moreover, our statistical security parameter is $\secpar_1=40$, meaning that with probability less or equal to $2^{-40}$, any adversary may learn something about the secret inputs.  To allow for better performance, we limit the fixed-point precision to $64$ bits, of which $h=32$ bits are allocated to the fractional part. This implies $p>2^{64}$, in order to avoid overflows. 
Usually higher precision would be necessary for the fault-free execution of our algorithms, but we enable lower precision by scaling numbers up or down in critical stages.
We let each prosumer run in one virtual machine. To simulate real-world conditions, the virtual machines communicate with each other via point-to-point TCP connections that are encrypted and authenticated by TLS.
Since RTTs higher than $\qty{10}{\milli\second}$ render our solution inefficient for most use cases, we provide benchmarks for $\{1,5,10\}\;\si{\ms}$ and simulate this latency between the virtual machines using Linux's traffic control (\textit{tc}).

\subsection{Benchmarks}
\label{ssec:benchmarks}

This section provides benchmarks for our privacy-preserving implementation of PFA via LU decomposition and GMRES. Out of the multitude of possibilities for privacy-preserving computation in the smart grid context, we propose a few setups that promise to be both realistic in terms of grid architecture and efficient in terms of runtime. We benchmark runtime for the following parameters of interest:
\begin{itemize}
    \item Offline vs. online phase: We give benchmarks for the full computation (meaning offline and online phase together) and also break out the online phase.
    \item Round Trip Time.
    \item Security model, meaning whether we assume semi-honest or malicious behavior of prosumers and whether we require an honest majority or allow a dishonest majority of prosumers.
    \item Number of computing parties in the grid.
\end{itemize}

\Cref{tab:resultsoffline} presents benchmarks for the online phase of a secure PFA. Here, we assume that the correlated randomness has been generated between the parties beforehand, or has been distributed by a trusted dealer.
\Cref{tab:resultsonline} presents benchmarks of securely running a PFA, including the secure interactive generation of correlated randomness.

\begin{table}[h]
\caption{Runtime in seconds, online phase only ($^\star$ fastest and $^{\dagger}$ slowest)}
\label{tab:resultsoffline}
\centering
\resizebox{\linewidth}{!}{
\begin{threeparttable}
{
\begin{tabularx}{\linewidth}{@{} c c@{\hspace{1em}} c c@{\hspace{1em}} c c@{\hspace{1em}} c @{}}\toprule
\multicolumn{1}{c}{} & \multicolumn{2}{c}{Semi-honest,}        & \multicolumn{2}{c}{Semi-honest, dis-}           & \multicolumn{2}{c}{Malicious, dis-} \\
\multicolumn{1}{c}{} & \multicolumn{2}{c}{honest majority$^{\mathrm{I}}$} & \multicolumn{2}{c}{honest majority$^{\mathrm{II}}$} & \multicolumn{2}{c}{honest majority$^{\mathrm{III}}$} \\
\multicolumn{1}{c}{RTT (\si{\ms})} & LU & GMRES & LU & GMRES & LU & GMRES \\
\midrule
\multicolumn{7}{l}{semi-urban grid with $n=39$} \\
1  & 72                & 111   & 62$^{\star}$ & 114              & 450 & 199 \\
5  & 2,227             & 3,549 & 280 & 535$\blind{^{\star}}$     & 2,701 & 1,073 \\
10 & 2,227 & 3,624 & 566 & 1,075$\blind{^{\star}}$     & 3,993$^{\dagger}$ & 1,672  \\
\midrule
\multicolumn{7}{l}{rural grid with $n=13$} \\
1  & 27$^{\star}$           & 41   & 56   & 92 & 52 & 85 \\
5  & 102$\blind{^{\star}}$  & 163  & 300  & 498 & 218 & 364 \\
10 & 197$\blind{^{\star}}$  & 316  & 566  & 935$^{\dagger}$ & 420 & 700\\
\bottomrule
\end{tabularx}}
\begin{tablenotes}
\vspace{.5mm}
    \item[] $^{\mathrm{I}}$Based on BGW \cite{AsharovL17} (``Shamir''). %
    $^{\mathrm{II}}$Semi-honest version of \cite{KellerOS16} (``Semi''). %
    $^{\mathrm{III}}$Based on \cite{CramerDESX18} (``SPD$\Z_{2^k}$'').
   \end{tablenotes}
\end{threeparttable}
}
\end{table}

\begin{table}[h]
\caption{Runtime in seconds, offline and online phase combined 
}
\label{tab:resultsonline}
\centering
\resizebox{\linewidth}{!}{
\begin{threeparttable}
{
\begin{tabularx}{\linewidth}{@{} c c@{\hspace{1em}} c c@{\hspace{1em}} c c@{\hspace{1em}} c @{}}\toprule
\multicolumn{1}{c}{} & \multicolumn{2}{c}{Semi-honest,}  & \multicolumn{2}{c}{Semi-honest, dis-}  & \multicolumn{2}{c}{Malicious,} \\
\multicolumn{1}{c}{} & \multicolumn{2}{c}{honest majority$^{\mathrm{I}}$} & \multicolumn{2}{c}{honest majority$^{\mathrm{II}}$} & \multicolumn{2}{c}{honest majority$^{\mathrm{III}}$} \\
\multicolumn{1}{c}{RTT (\si{\ms})}\!\!\!\!\!\!\!\!\! & LU & GMRES\!\!\!\! & LU & GMRES  & LU & GMRES \\
\midrule
\multicolumn{5}{l}{semi-urban grid with $n=39$} \\
1  & 2,584$^{\star}$ & 5,598 & \multicolumn{4}{c}{$>$24 hours} \\
5  & 54,880 & $>$24 hours\!\!\! &\multicolumn{4}{c}{$>$24 hours} \\
\midrule
\multicolumn{5}{l}{rural grid with $n=13$} \\
1  & 234   & 232$^{\star}$    &  1,994   & 1,570   & 11,774$\blind{^\dagger}$   & 10,445 \\
5  & 387   & 424$\blind{^{\star}}$  & 6,023 & 4,655 & 18,358$\blind{^\dagger}$   & 16,388 \\
10 & 584 & 672$\blind{^{\star}}$ &  10,800 & 8,291 & 24,943$^{\dagger}$\!\! \!       & 22,356 \\
\bottomrule
\end{tabularx}}
\begin{tablenotes}
\vspace{.5mm}
    \item[] $^{\mathrm{I}}$Based on \cite{GoyalLOPS21} (``Atlas'') in the semi-urban case and on BGW \cite{AsharovL17} (``Shamir'') in the rural case.
    $^{\mathrm{II}}$ Based on \cite{Cramer2001multiparty} and \cite[Appendix B.1]{Keller022} using LWE (``Temi'').
    $^{\mathrm{III}}$ Based on \cite{ChidaHIKGLN23} (``Sy-Shamir'').
   \end{tablenotes}
\end{threeparttable}}
\end{table}

We can see from the benchmark data that grid size and network latency have a significant
impact on the runtime of our privacy-preserving PFA. If we assume that the prosumers are connected in a LAN via Ethernet, where RTTs of $\qty{1}{\ms}$ are realistic, runtimes of under 30 seconds (for the online phase) and four minutes (for the entire computation, i.e., preprocessing and online phase) are possible in small grids with 13 prosumers. If prosumers are expected to deviate from the protocol, security against malicious behavior is required. Here, runtimes of one minute and under three hours are possible in the online phase and for the entire computation, respectively.
As we increase latency and grid size, our measured runtime increases so that in a grid with 39 prosumers online computation is finished in approximately
two minutes, but preprocessing and online phase together take at least 70 minutes.
As described in more detail below, the expensive preprocessing phase can, however, be omitted under certain assumptions.

\subsection{Discussion}
Many smart meters today communicate via unoptimized Ethernet, broadband over power lines (BPL) or LTE connections characterized by relatively high RTTs in the range of $\qty{20}{\milli\second}$ to $\qty{150}{\milli\second}$, which would render our solution impractical. Assuming an optimized setup, however, RTTs of $1$ ms for LAN-based connections, and $5$ ms for WAN-based connections between smart meters would be possible. 
For a holistic assessment of practicality, one should also consider hardware constraints: We used a high-performance server for our benchmarks, on which 13 resp.~39 virtual machines (each representing one prosumer) shared 16 \qty{3.5}{\giga\hertz} \textit{Ryzen 9} cores. Nevertheless, the use of more resource-constrained processors usually integrated into smart meters should not be a problem, at least in the online phase of protocols secure against semi-honest adversaries. This is because in these cases, even with a short RTT of \qty{1}{\milli\second}, time spent on communication dominates CPU time by a factor of ten. With longer RTTs such as \qty{10}{\milli\second}, time spent on communication can dominate CPU time by a factor of 100.
Moreover, RAM load was roughly \qty{0.5}{\giga\byte} to \qty{0.75}{\giga\byte} per prosumer during the online phase. While this is slightly more than typically available in smart meters, it should be noted that our protocols can still be optimized for memory usage. Moreover, the \qty{0.5}{\giga\byte} includes system overhead, which is likely going to be much lower in the operating systems deployed in smart meters. 
To allow on-device preprocessing and protocols secure in the malicious model, however, hardware improvements to smart meters might be necessary.
In addition, even given sufficiently powerful hardware, generation of correlated randomness might take too long to complete offline, especially in use cases with only a small idle period between computations.

For these use cases, it may make sense to outsource the generation and distribution of correlated randomness to a trusted dealer. As long as this dealer is oblivious to the communication transmitted during the online phase, it will learn nothing about the secret inputs. In extension, this means that we also have to assume that none of the prosumers is colluding with the dealer. 
We can reduce the trust required in the dealer by using a setup where the correlated randomness is generated interactively by two non-colluding dealers (e.g., the grid operator and a governmental regulator) such that both dealers only ever see partial shares ('shares of shares') of the correlated randomness assigned to a certain prosumer. Both dealers distribute their partial shares to the respective prosumers, who can then reconstruct their full shares of the correlated randomness from them. This would allow outsourcing of the expensive preprocessing phase.

\section{Conclusion and Outlook}
\label{sec:conclusion}
In this work, we investigated the practicality
of SMPC techniques for privacy-preserving PFA in the context of smart grids. To this end, we devised a 
power flow 
solution algorithm implemented a 
privacy-preserving
PFA, and provided runtime benchmarks for various SMPC protocols, grid parameters, and threat models. Further, we optimized the efficiency of our implementation through various means such as batching of communication rounds and enabling lower precision by dynamic rescaling.

Our method is particularly suited for market-based and preventive smart grid applications where prosumer data is used. In these use cases, PFA is performed as part of a grid state forecast to detect grid congestions at an early stage. 
To solve these grid congestions, market-based solutions like local flexibility markets \cite{Amicarelli17}\cite{Esmat18}, preventive control concepts, or dynamic grid charges are possible. Since the lead times of these solutions leave between fifteen minutes and one day for running PFA, our privacy-preserving solution is well-suited here. On the other hand, we have shorter lead times of ten seconds to one minute in real-time smart grid control systems, and currently, our implementation can only accommodate these runtimes for small grids with a short RTT.

Real-time smart grid applications might hence be addressed in future research. Due to the availability of better tooling, we focused on SMPC via secret sharing in this work. However, it would also be possible to achieve SMPC by means of \textit{threshold fully homomorphic encryption} where the secret key is distributed among many parties such that the input of all of those parties is required for decryption. In this setting, all parties would encrypt their secret inputs and send the encryptions to a semi-honest server that will perform the computation. Once the function has been evaluated, all parties get the output ciphertext and partake in the decryption. The advantages of this approach are that the communication load is low and that the computational load can be securely outsourced to a high-performance server, potentially allowing for shorter runtimes.

Another direction for future research is to explore alternative methods for solving the power flow problem. 
In this context, the holomorphic embedding load flow method~\cite{Trias2012} stands out due to its guaranteed convergence to the operable correct solution. Essentially, this is realized by replacing complex voltage variables by a holomorphic function, which results in a solution algorithm that may be well aligned with SMPC protocols.

\bibliographystyle{IEEEtran}
\bibliography{references} 

\appendix

\subsection{Estimation of the admittance matrix}
\label{app:admittance}
The admittance matrix $\Yb$, determined by $\Gb$ and $\Bb$, can be robustly estimated based on the measurements 
$\{\vb_{R,\kappa},\vb_{I,\kappa},$ $\ib_{R,\kappa},\ib_{I,\kappa}\}_{\kappa=0}^T$
where  $T\in\N$ is sufficiently large. To this end, we use 
$\xb_{\kappa}^\top=(\vb_{R,\kappa}^\top,\vb_{I,\kappa}^\top)$
and define 
$\wb_{\kappa}^\top = (\ib_{R,\kappa}^\top,\ib_{I,\kappa}^{\top})$
such that 
$\wb_{\kappa} = \Wb \xb_\kappa$
with
\begin{equation*}
    \Wb = \begin{pmatrix} \Gb & -\Bb \\ \Bb  & \blind{-}\Gb \end{pmatrix}.
\end{equation*}
Then, the optimizer 
$\argmin_{\Wb}\sum_{\kappa=0}^T \| \Wb \xb_\kappa -\wb_\kappa\|_{2}^2$
is 
\begin{align}
\label{eq:Wopt}
\vect{\Wb^\star} =
\begin{pmatrix}
        \xb_0^\top \otimes \mathbb{I} \\
        \vdots \\
        \xb_T^\top \otimes \mathbb{I} 
    \end{pmatrix}^{+}
    \begin{pmatrix}
        \wb_0 \\
        \vdots \\
        \wb_T
    \end{pmatrix},
\end{align}
where $(\cdot)^+$ denotes the Moore-Penrose inverse and $\otimes$ is the Kronecker product.
In case only measurements $\{\vb_{R,\kappa},\vb_{I,\kappa},\pb_{\kappa},\qb_{\kappa}\}_{\kappa=0}^T$
are available, one can build on \eqref{eq:powerandvoltage}. 
In comparison to before, we define 
$\tilde{\wb}_\kappa^\top = (\pb_\kappa^\top,\qb_\kappa^\top)$
and 
$$
    \Vb(\xb)=\begin{pmatrix}
        \diag(\vb_R) & \!\!\! \blind{-}\diag(\vb_I) \\
        \diag(\vb_I) & \!\!\! -\diag(\vb_R)
    \end{pmatrix}
$$
such that \eqref{eq:powerandvoltage} becomes 
$\tilde{\wb}_{\kappa} = \Vb(\xb_\kappa)\Wb\xb_\kappa$.
Then, $\Wb^\star$ is obtained from \eqref{eq:Wopt} by replacing $\mathbb{I}$ with 
$\Vb(\xb_\kappa)$ and $\wb_\kappa$ with $\tilde{\wb}_{\kappa}$.
Note that \eqref{eq:Wopt} can be solved privately (without revealing any of the measurements) using the methods presented above to compute, e.g., a
singular value decomposition. 
Furthermore, there are no hard restrictions on the computation time for $\Wb^\star$.

\subsection{Omitting pivoting}
\label{app:pivot}

It remains to clarify, why pivoting can often be omitted in the solution of \eqref{eq:newton} via an LU decomposition. To this end, we note that pivoting is necessary to avoid a division by $0$, which occurs if an element in $\diag(\Jb(\xb))$ is $0$. There, $\diag(\Jb(\xb))$ denotes setting all but the main diagonal elements of $\Jb(\xb)$ to zero. In other words, as long as the elements in $\diag(\Jb(\xb))$ are sufficiently far away from $0$, pivoting is not required. Our ansatz is now to check a priori that this is fulfilled for $\xb\in\R^{2 (N-1)}$ based on $\Gb$ and $\Bb$. By inspection of \eqref{eq:jacobian} one can see that $\diag(\Jb)$ depends on $\diag(\Hb_3\pm \Hb_1)$
which can be rearranged into
\begin{align*}
\begin{pmatrix} \Gb  + \diag(\Gb)  &  -\Bb+\diag(\Bb)
\\
\Gb - \diag(\Gb)  & -\Bb-\diag(\Bb)
\end{pmatrix}
\begin{pmatrix} \vb_R\\ \vb_I \end{pmatrix}
=\Kb \xb.
\end{align*}
Then, $\Kb\xb\neq \zerob$ without $\xb=\zerob$ requires an empty nullspace of $\Kb$ which is the case if $\det(\Kb)\neq 0$.
Furthermore, 
one can efficiently compute a bound from zero for
the elements in $\diag(\Jb)$
by, e.g., using a binary search for the smallest positive $\varepsilon_i$ for all $i\in\{1,\dots,2N\}$ such that $\sigma^\star > 0$ resulting from
\begin{align*}
    \min_{\sigma,\xb} \sigma \;\; \text{s.t.} \;\; -\sigma- \varepsilon_i \leq \Kb_{i,:}\xb\leq \varepsilon_i+\sigma, \;\; 0 \leq \sigma,
\end{align*}
is fulfilled. 
Clearly, $ \Kb_{i,:} \xb \notin [-\varepsilon_i,\varepsilon_i]$ as long as $\sigma^\star>0$, because otherwise $\sigma=0$ would have been optimal.
Finally, the bound is $\min\{\varepsilon_1,\varepsilon_2,\ldots,\varepsilon_{2N}\}$. Still, if the aforementioned conditions fail, one could add a small disturbance $\delta$ as in $\Jb+\delta \mathbb{I}$, where $\mathbb{I}$ is a conformal identity matrix. Then, a relative error $e$ for $\|\Delta \xb\|_2$ is achieved when $\delta = e/\|\Jb\|_2$ (see \cite[Chapter 2]{demmel1997applied}).

\end{document}